\documentclass[structureabstract]{aa}  
\usepackage{graphicx}
\usepackage{txfonts}
\begin{document}
\authorrunning{Montes et al.}
\titlerunning{A Model for the thermal radio emission from radiative shocks
 in binaries}

   \title{A model for the thermal radio-continuum emission from radiative
 shocks in colliding stellar winds}


   \author{G. Montes,
          \inst{1}
          R.F. Gonz\'alez,
          \inst{2}
          J.Cant\' o,
          \inst{3}
          M.A. P\'erez-Torres,
          \inst{1}
          \and
          A. Alberdi
          \inst{1}         
          }

   \institute{Instituto de Astrof\'\i sica de Andaluc\'\i a (IAA), CSIC,
              Camino Bajo de Huetor 50, E-18006 Granada, Spain\\
              \email{gmontes@iaa.es}
         \and
             Centro de Radioastronom\'\i a y Astrof\'\i sica, UNAM, Mexico\\
             \email{rf.gonzalez@crya.unam.mx}
         \and
         Instituto de Astronomia (IA), UNAM, Mexico
             }

   \date{Received  October, 2010; accepted April, 2011}


  \abstract
   {In massive-star binary systems, the interaction of the strong stellar
winds results in a wind collision region  between the stars, which is limited by two shock
fronts.  Besides the nonthermal emission
resulting from the shock acceleration, these shocks emit thermal (free-free)
radiation detectable at radio frequencies that increase the expected
emission from the stellar winds. Observations and theoretical studies
of these sources show that the shocked gas is an important, but not
dominant, contributor to the total emission in wide binary systems,
while it plays a very substantial role in close binaries.}
   {The interaction of two isotropic stellar winds is studied in order
to calculate the free-free emission from the wind collision region.
The effects of the binary separation and the wind momentum ratio on the
emission from the wind-wind interaction region are investigated.}
   {We developed a semi-analytical model for calculating the thermal
emission from colliding stellar winds. Assuming radiative shocks for
the compressed layer, which are expected in close binaries,
we obtained the emission measure of the thin shell. Then, we computed the
total optical depth along each line of sight to obtain the
emission from the whole configuration.}
   {Here, we present predictions of the free-free emission at radio
frequencies from analytic, radiative shock models in colliding wind
binaries. It is shown that the emission from the wind collision region
mainly arises from the optically thick region of the compressed layer
and scales as $\sim D^{4/5}$, where $D$ is the binary separation. The
predicted flux density $S_\nu$ from the wind collision region
becomes more important as the frequency $\nu$ increases, showing
higher spectral indices than the expected 0.6 value ($S_\nu \propto
\nu^\alpha$, where $ \alpha =0.6$) from the unshocked winds. We also
investigate the emission from short-period WR+O systems calculated with
our analytic formulation. In particular, we apply the model to the binary
systems WR 98 and WR 113 and compare our results with the observations.
Our theoretical results are in good agreement with the observed thermal
spectra from these sources.}
   {}


   \maketitle
%

\section{Introduction}

Stellar winds from hot massive stars, OB and Wolf-Rayet (WR) type stars, emit
free-free thermal emission  detectable at radio frequencies. Stars with
spherically symmetric, isothermal, and stationary outflows are predicted to produce
radio spectra with a characteristic frequency dependence of $S_{\nu} \propto
\nu^{0.6}$ (see, Panagia \& Felli, 1975; Wright \& Barlow 1975). The spectral
index $\alpha = 0.6$ results from the radial dependence of the electron density,
$n \propto r^{-2}$.
These authors also show that variations in this electron density behavior
results in free-free spectra with a frequency dependence different from the
0.6 value. Likewise, Leitherer $\&$ Robert (1991) discuss several
mechanisms that may cause a deviation from a 0.6 spectral index focusing on
changes in the ionization structure and velocity gradients in the wind. Such
effects may be present and produce small variations in the observed slope of the
radio spectrum ($0.6 < \alpha < 0.7$). In addition, Gonz\'alez $\&$ Cant\'o
(2008) showed that variabilities in the wind parameters (such as velocity
and mass loss rate) at injection produce internal shocks that generate
thermal continuum radiation detectable at radio wavelengths. Their predicted
spectral indices clearly deviate from the expected 0.6 value of the standard
model.

In binary systems, the stellar winds of the components must collide,
resulting in the formation of a two-shock wave structure between the stars
(see, for instance, Eichler \& Usov 1993). The wind collision region (WCR)
emits both thermal and nonthermal radiation (e.g. Pittard et al. 2006).
The thermal component is readily explained as free-free emission, while
the nonthermal component is thought to be synchrotron radiation
arising from electrons accelerated at the shocks bounding the WCR
in the wind-wind interaction zone (see, for instance, Williams et al. 1990, 1997). 
In wide binaries, observations at radio frequencies show negative spectral
indices that suggest that the contribution of nonthermal emission
dominates the total spectrum.
On the other hand, in close systems, the synchrotron radiation must be
produced within the optically thick region of the unshocked stellar winds,
and then it is expected to be highly attenuated by free-free absorption
(e.g., Chapman et al. 1999; Monnier et al. 2002).
The first quantitative investigation of the effect of binarity on thermal
radio emission was performed by Stevens (1995), who shows that the
presence of a companion with a strong wind increases the expected thermal
radio emission as compared to a single star with the same wind parameters
as the primary. In wide systems, the shocked gas is an important but not
dominant contributor, while in close systems it plays a very substantial
role in the excess radio emission. 
In addition, Kenny $\&$ Taylor (2005) present colliding wind models for
symbiotic star systems by assuming good mixing of the shocked material from
both winds. They performed three-dimensional simulations of
bremsstrahlung radio images, and adopted fully ionized colliding winds.
They found spectral differences associated with the viewing angle, which
imply that the flux and spectral index vary with orbital phase. 

Pittard et al. (2006) carried out numerical models for computing the radio 
continuum contribution from adiabatic shocks in a colliding wind binary,
and showed that the WCR clearly impacts on the thermal spectrum.
They also found that the hot gas within the WCR remains optically thin,
with an intrinsic emission of spectral index $\alpha_{WCR}\sim -0.1$.
These authors investigated how the thermal
flux from the WCR varies with binary separation, and found that its
free-free emission scales as $D^{-1}$, where $D$ is the binary separation.
Consequently, they pointed out that a composite-like spectrum (which may
suggest nonthermal emission) can result entirely from thermal processes.
More recently, Pittard (2010) developed 3D hydrodynamical models for computing
the thermal radio to submillimeter emission from radiatively colliding wind
binaries (in O+O star systems; see also Pittard 2009). In these models, the
flux density and the spectrum as a function of orbital phase and orientation
to the observer are investigated. 
They computed flux variations with orbital phase, which are caused by 
changes in the relative position  of the emitting components with the
orbital motion.
In particular, they investigated an eccentric
system where the physical properties (such as density and temperature) of
the WCR change along the orbit, which is optically thick (highly radiative) at
periastron but optically thin (adiabatic) at apastron.

Observations of stellar winds at radio wavelengths from some hot stars
(see, for instance, Leitherer $\&$ Robert 1991; Altenhoff, Thum, $\&$
Wendker 1994; Nugis, Crowther, $\&$ Willis 1998) show flux densities and
spectral indices that differ from the expected value for the uniformly
expanding wind model. On the other hand, Montes et al. (2009) presented
multi-frequency radio observations of a sample of several WR stars, and
discussed the possible scenarios to explain the nature of their emission.
From these data, they found evidence for sources with thermal (free-free
thermal emission), nonthermal (dominant synchrotron emission), and
composite (thermal+nonthermal) spectra. Nevertheless, close binaries
classified as thermal+nonthermal sources cannot be completely ruled out
as thermal sources (as pointed out by Pittard et al. 2006), and theoretical
models are required to determine whether a composite spectrum can be
reproduced entirely by thermal emission.

Cant\'o, Raga, $\&$ Wilkin (1996) developed a formalism based on linear
and angular momentum conservation for solving steady thin-shell problems,
which is applicable to the interaction of non-accelerated flows. In particular,
these authors found analytic solutions to the case of two colliding isotropic
stellar winds. Assuming that the postshock fluid is well mixed across the
contact discontinuity, they obtained the shape of the thin shell, its mass
surface density, and the velocity along the layer. Cant\'o, Raga,
$\&$ Gonz\'alez (2005) also present a model for calculating the emission measure
from thin-shell flow problems. In particular, these authors applied the
results obtained by Wilkin (1996; 2000) and Cant\'o, Raga, $\&$ Wilkin (1996)
for a stellar wind interacting with an impinging ambient flow with a density
stratification. From their model, simple predictions of the free-free emission
can be made. 

In this work, we study the case of the interaction of two
spherically symmetric stellar winds. We present a formulation for calculating
the free-free emission from the wind-wind interaction region, with which
the total thermal spectrum of colliding wind binaries can be obtained.
This paper is organized as follows. In $\S$ 2, we describe the
model. In $\S$ 3, we show the results of our model
for the predicted thermal fluxes and the corresponding spectral
indices for massive binary systems. A comparison with observations
of this kind of sources are presented in $\S$ 4. Finally, we give
our conclusions in $\S$ 5.

\section{The analytical model}

We consider a binary system that contains colliding isotropic
stellar outflows. This situation is shown in a schematic way in Figure 1.
We assume two ionized winds that move radially away from the stars
separated by a distance $D$. The stellar winds are generally accelerated
to hypersonic speeds and reach a good fraction of their terminal velocities
in a few stellar radii. In our model, we assume that the winds are ejected
with terminal speeds. The interaction of these two outflows gives rise to
the wind collision region (WCR) that is, in principle, composed of two
shocks separated by a contact discontinuity (Luo et al. 1990). If the shocks
are radiative (post-shock cooling becomes very efficient leading to a large
compression), the shock fronts collapse onto a thin shell and the width of
the WCR can be neglected.

Let $\dot m_1$ and $v_1$ be the mass loss rate and the velocity of the wind
source at the origin of the spherical coordinate system $(R,\theta,\phi)$,
and $\dot m_2$ and $v_2$ the corresponding quantities for the wind from the
source located at a distance $D$. Using the geometric relation
\begin{eqnarray}
R(\theta)= D \,\mbox{sin}\, \theta_1 \, \mbox{csc}\, (\theta + \theta_1)\,,
\label{eq1}
\end{eqnarray}
\noindent
where $R(\theta)$ is the radius of the layer, Cant\'o, Raga, $\&$ Wilkin (1996)
obtained the exact (implicit) analytic solution
\begin{eqnarray}
\theta_1\,\mbox{cot}\, \theta_1 = 1 + \beta\, (\theta \,\mbox{ctg}\,
\theta - 1)\,,
\label{eq2}
\end{eqnarray}
where the dimensionless parameter $\beta = (\dot m_1\, v_1) / (\dot m_2\, v_2$)
is the wind momentum ratio. For a given $\theta$, equation (\ref{eq2}) can
be solved numerically for $\theta_1$, and the radius is then obtained from
equation (\ref{eq1}). For the case of an isotropic stellar wind, the flow is
axisymmetric, and therefore the spherical radius $R$ does not depend on the
azimuthal angle $\phi$.  In addition, the stagnation point radius,
$R_0= \beta^{1/2} D/(1 + \beta^{1/2})$ is obtained from the ram-pressure
balance condition (Cant\'o et al. 1996).

\begin{figure}
\centering
\includegraphics[width=8cm]{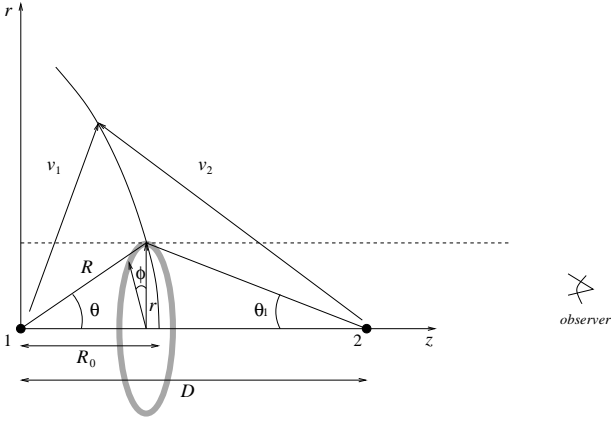}
\caption{Schematic diagram showing the interaction of two spherical
winds that move radially away from the stars. Source 1 of the wind
with velocity $v_1$ is located at the origin of the coordinate
system, and source 2 of the wind with velocity $v_2$ at a distance
$D$ along the symmetry axis. The shape of the layer where the winds
collide (at $R_0$ along the symmetry axis) is given by the curve
$R(\theta$). The angles $\theta$ and $\theta_1$ are measured from the
positions of the stars to the intersection point between a line of sight
(with impact parameter $r$) and the layer. We assume azimuthal symmetry
(no dependence on angle $\phi$). The observer is located in the
orbital plane along the z-axis of symmetry.}
\label{f1}
\end{figure}

The pressure within  the shell just behind each shock can be written in terms
of the preshock density, and the velocity component normal to the shell.
The normal vector to the surface of the shell can be obtained from the
gradient of the  function $F=r'-R(\theta)$, where $r'$ is the radial
coordinate, and the bow shock is described by $F=0$. In this way, $\hat{n}=
\nabla F / \mid\nabla F\mid$, where $\nabla$ denotes the vector differential
operator gradient in spherical coordinates (see Wilkin 1997). For isotropic
outflows $\partial R/\partial \phi = 0$ (see eq.(\ref{eq1})), and then
\begin{eqnarray}
\hat{n}= {{\hat{R} - 1/R \,(\partial R/ \partial \theta) \, \hat{\theta}}
  \over{\sqrt{(1 + 1/R^2 \,(\partial R/ \partial \theta)^2)}}} \,,
\label{eq3}
\end{eqnarray}
where $\hat{R}$ and $\hat{\theta}$ are unit vectors measured along the directions
$R$ and $\theta$, respectively.

On the other hand, the velocity vectors of the stellar
winds are given by
\begin{eqnarray}
\vec{v_1}= v_1 \,\hat{R} \,,
\label{eq4} 
\end{eqnarray}
\noindent
and
\begin{eqnarray}
\vec{v_2}= - v_2 \,\,\mbox{cos}\,(\theta + \theta_1)\,\hat{R}
 + v_2\,\mbox{sin}\,(\theta + \theta_1) \,\hat{\theta} \,.
\label{eq5} 
\end{eqnarray}

From equations (\ref{eq3}), (\ref{eq4}) and (\ref{eq5}) it follows that the
normal component at every point $(R,\theta,\phi)$ of the pre-shock velocity
of the stellar winds are obtained by
\begin{eqnarray}
v_{1,n}=  v_{1}\,{{R}
 \over{\sqrt{R^2 + (\partial R/ \partial \theta)^2}}} \,,
\label{eq6}
\end{eqnarray}
\begin{eqnarray}
v_{2,n}=  -v_{2}\, {{R\,\mbox{cos}\,(\theta + \theta_1) +
 (\partial R/ \partial \theta})\, \mbox{sin}\,(\theta + \theta_1)
 \over{\sqrt{R^2 + (\partial R/ \partial \theta)^2}}} \,.
\label{eq7}
\end{eqnarray}

As a result, the pressure within the thin shell just behind each shock front
can be written as
\begin{eqnarray}
P_{1}= \rho_{1,0}\, v_{1,0}^{2} \, f_{w,1} \,(\theta,\theta_1) \,,
\label{eq8}
\end{eqnarray}
and
\begin{eqnarray}
P_{2}= \rho_{1,0}\, v_{1,0}^{2} \, f_{w,2} \,(\theta,\theta_1) \,,
\label{eq9}
\end{eqnarray}
\noindent
where we have used the wind parameters $\rho_{1,0}$ and $v_{1,0}^{2}$
at the stagnation point $R_0$ to nondimensionalize the equations, and
\begin{eqnarray}
f_{w,1}\,(\theta,\theta_1)=  \biggl({{\rho_{1}\, v_{1}^{2}}\over{\rho_{1,0}\,
 v_{1,0}^{2}}}\biggr)\, {{R^2}\over{R^2 + (\partial R/ \partial \theta)^2}}, 
\label{eq10}
\end{eqnarray}
and
\begin{eqnarray}
f_{w,2}\,(\theta,\theta_1)= \biggl({{\rho_{2}\, v_{2}^{2}}
\over{\rho_{1,0}\, v_{1,0}^{2}}}\biggr)\,\times
\nonumber
\end{eqnarray}
\begin{eqnarray}
 {{{[R\,\mbox{cos}\,(\theta + \theta_1) + (\partial R/ \partial \theta)\,
 \mbox{sin}\,(\theta + \theta_1)]}^2} \over{R^2 + (\partial R/ \partial
 \theta)^2}}\,,
\label{eq11}
\end{eqnarray}
\noindent
where $\rho_{1}$ and $\rho_{2}$ are the preshock densities of the stellar winds.

\subsection{The emission measure of the thin shell}

Let $l$ be a longitude-coordinate measured inwards from the wind shock of
source 2 and perpendicular to the thin shell (as shown in Fig. 2). Then, the
shock fronts are located at $l= 0$ and $l= h$, where $h$ is the position-dependent
thickness of the shell. We assume that the post-shock gas is well mixed, so that
it flows along the shell at a velocity $v$, which is independent of $l$. If the
whole flow is photoionized and approximated as an isothermal flow, the pressure as
a function of the parameter $l$ can be calculated from the hydrostatic equation
\begin{eqnarray}
{{dP}\over{dl}}= - \rho \,g \,,
\label{eq12}
\end{eqnarray}
\noindent
where $P= \rho c_s^2$ (with $c_s$ being the isothermal sound speed), and
the centrifugal acceleration $g= v^2/R_c$, with $R_c$ the radius of curvature
of the thin shell. Integrating equation (\ref{eq12}) with the boundary
condition $P_2= P(l=0)$, we obtain
\begin{eqnarray}
P(l)=  \rho(l)\,c_s^2 = P_2\,\mbox{e}^{-l/H},
\label{eq13}
\end{eqnarray}
\noindent
where $H= c_s^2/g$ is the scale height parameter of pressure (or density).
It follows from equation (\ref{eq13}) that the surface density is given by
\begin{eqnarray}
\sigma= \int_0^h \rho(l)\,dl = {{H}\over{c_s^2}}\,(P_2 - P_1).
\label{eq14}
\end{eqnarray}
\noindent
Using equations (\ref{eq13}) and (\ref{eq14}), we can now calculate the
emission measure of the thin shell,
\begin{eqnarray}
EM= \int_0^h (\rho/\bar m)^2 \,dl = {{\sigma} 
 \over{2 \bar{m}^2 c_s^2}}\,(P_2 + P_1)\,,
\label{eq15}
\end{eqnarray}
\noindent
where $\bar{m}=\mu m_p$ is the average mass per particle of the photoionized
gas, $\mu$ the mean atomic weight per electron, and $m_p$ the mass of
the proton.

Equation (\ref{eq15}) gives the emission measure of the thin shell in terms
of its surface density. Next, we must obtain the surface density as a function
of position on the shell. We first note that the shocked gas can be described
as a flow of streamlines with constant azimuthal angle $\phi$. The mixing
of the two shocked winds is assumed to be instantaneous, so that the flow will
have a unique velocity $v$ at any location with in the shell. The mass (per
unit time) $\Delta \dot M$ flowing along a ring of radius $r$ of the
layer is then given by
\begin{eqnarray}
\Delta \dot M= 2 \pi\, R \,\mbox{sin}\,\theta\, \sigma v\,,
\label{eq16}
\end{eqnarray}
\noindent
which must be equal to the mass injection rate (into the same solid angle)
of the stellar winds, that is,
\begin{eqnarray}
\Delta \dot M= {{\dot m_1}\over{2}}\, f_m(\theta, \theta_1)\,,
\label{eq17}
\end{eqnarray}
\noindent
with
\begin{eqnarray}
f_m(\theta, \theta_1)= 1 - \mbox{cos}\,\theta + 
 {{v_1}\over{\beta\,v_2}} (1 - \mbox{cos}\,\theta_1)\,.
\label{eq18}
\end{eqnarray}

\begin{figure}[t!]
\centering
\includegraphics[width=7cm]{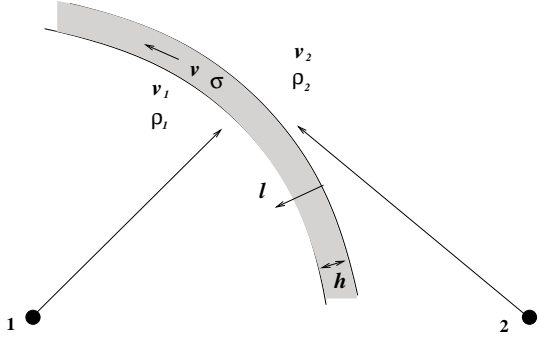}
\caption{Schematic diagram showing a thin shell (limited by two shock
fronts) resulting from the interaction of two stellar winds. The shocked
gas flows along the shell at a velocity $v$ and has a surface density
$\sigma$. The coordinate $l$ is measured inwards from the wind shock
of source 2, normal to the locus of the thin shell.}
\label{f2}
\end{figure}

Based on considerations of linear momentum conservation, it can be shown
that the velocity along the shell (see also Cant\'o, Raga, $\&$ Wilkin
1996; Cant\'o, Raga, $\&$ Gonz\'alez 2005) can be written as
\begin{eqnarray}
v= v_1 {{\sqrt{f_r^2\,(\theta, \theta_1) + f_z^2\,(\theta, \theta_1)}}
\over{f_m(\theta, \theta_1)}}\,,
\label{eq19}
\end{eqnarray}
\noindent
where the functions $f_r\,(\theta, \theta_1)$ and $f_z\,(\theta, \theta_1)$
are deduced from $r$-momentum and $z$-momentum (being $r$ and $z$ the
directions of the cylindrical radius and the symmetry axis, respectively),
and are given by
\begin{eqnarray}
f_r(\theta, \theta_1)= {{1}\over{2}}\,\biggl[\theta - \mbox{sin}\,\theta\,
 \mbox{cos}\,\theta +  {{1}\over{\beta}}\, (\theta_1 -
 \mbox{sin}\,\theta_1\,\mbox{cos}\,\theta_1)\biggr]\,,
\label{eq20}
\end{eqnarray}
\noindent
and
\begin{eqnarray}
f_z(\theta, \theta_1)= {{1}\over{2}}\,\biggl[ \mbox{sin}^2\,\theta\,
  -  {{1}\over{\beta}}\,\mbox{sin}^2\,\theta_1\ \biggr]\,.
\label{eq21}
\end{eqnarray}

Substitution of equations (\ref{eq17})-(\ref{eq21}) into equation (\ref{eq16})
gives the surface density,
\begin{eqnarray}
\sigma= \sigma_0 \,f_{\sigma}(\theta, \theta_1)\,,
\label{eq22}
\end{eqnarray}
\noindent
where $\sigma_0= \dot m_1/ (2\pi \beta D v_1)$ and
\begin{eqnarray}
f_{\sigma}(\theta, \theta_1)= {{\beta}\over{2}}\,\mbox{sin}\,(\theta + \theta_1)
 \,\mbox{csc}\,\theta_1\, \mbox{csc}\,\theta\,
 {{f_m^2(\theta, \theta_1)}\over{\sqrt{f_r^2(\theta, \theta_1)
 + f_z^2(\theta, \theta_1)}}}\,.
\label{eq23}
\end{eqnarray}

Finally, it follows from equation (\ref{eq15}) that the emission measure
of the thin shell (as a function of position) is given by
\begin{eqnarray}
EM(\theta, \theta_1)= EM_0\,f_{\sigma}(\theta, \theta_1)
  \,[f_{w,1}\,(\theta,\theta_1) + f_{w,2}\,(\theta,\theta_1)]\,,
\label{eq24}
\end{eqnarray}
\noindent
with $EM_0= \sigma_0\, \rho_{1,0}\, v_{1,0}^{2}/(2 \bar{m}^2 c_s^2)$.

\subsection{Predicted thermal radio emission from a binary system}

To calculate the radio-continuum emission from the whole system,
it is necessary to compute the total optical depth along each line of sight,
which will have the contribution from the stellar winds and also from the thin
shell. Then we estimate the intensity emerging from each direction and
calculate the flux by integrating the intensity over the solid angle.

Given the emission measure $EM(\theta, \theta_1)$ of the shocked
layer (eq.(\ref{eq24})), the optical depth perpendicular to the WCR
is calculated by $\tau_{WCR,\perp}(\theta, \theta_1)= EM(\theta, \theta_1)\,
\chi(\nu)$, where $\chi(\nu)=$8.436 $\times$ 10$^{-7} \, \nu^{-2.1}$ with
 the frequency $\nu$ in Hz (for an electron temperature $T_e=$ 10$^4$K).
 By defining a critical frequency, such that
\begin{eqnarray}
\biggl({{\nu}\over{\nu_c}}\biggr)^{-2.1}= \biggl({{\dot m_1}
 \over{4 \pi \bar{m} v^2_{1,0}}}\biggr)^2\,{{\chi(\nu)}\over{R_0^3}}\,,
\label{eq25}
\end{eqnarray}
\noindent
it can be shown from equation (\ref{eq24}) that
\begin{eqnarray}
\tau_{WCR,\perp}(\theta, \theta_1)= \biggl({{\nu}\over{\nu_c}}\biggr)^{-2.1}
\,\biggl({{v_{1,0}}\over{c_s}}\biggr)^2 \,\times
\nonumber
\end{eqnarray}
\begin{eqnarray}
\biggl(\,{{1}\over{\beta \tilde D}}\biggr)\, 
 f_{\sigma}(\theta, \theta_1) \,[f_{w,1}\,(\theta,\theta_1) +
 f_{w,2}\,(\theta,\theta_1)]\,,
\label{eq26}
\end{eqnarray}
\noindent
where $\tilde D= D/ R_0$. For a line of sight intersecting the thin
shell at an angle $\mbox{cos}^{-1}\,(\hat{z} \cdot \hat{n})$ from the
normal (being $\hat{z}= \mbox{cos}\,\theta \,\hat{R} - \mbox{sin}\,\theta
\,\hat{\theta}$), the optical depth is then given by
\begin{eqnarray}
\tau_{WCR}(\theta, \theta_1)= \tau_{WCR,\perp}(\theta, \theta_1)
{{\sqrt{\tilde{R}^2 + (\partial \tilde{R}/ \partial \theta)^2}}
\over{\tilde{R}\,\mbox{cos}\,\theta + \sin \theta\,
(\partial \tilde{R}/ \partial \theta)}}\,,
\label{eq27}
\end{eqnarray}
\noindent
where $\tilde{R}$ is in units of $R_0$.

Let us now calculate the contribution of the unshocked stellar winds
to the optical depth. Consider a line of sight that intersects the thin
shell at a point $(\tilde{r}\, \mbox{ctg}\, \theta, \tilde{r})$, as shown
in Figure 1. According to Panagia $\&$ Felli (1975) and Wright $\&$ Barlow
(1975), the optical depth along the line of sight of the wind source
located at $z= 0$ is obtained by
\begin{eqnarray}
\tau_{w,1} (\theta)=
 \int_{-\infty}^{\tilde{r}\, \mbox{ctg}\, \theta} n_{w,1}^2(z)\,\chi(\nu)\,dz\,,
\label{eq28}
\end{eqnarray}
where
\begin{eqnarray}
n_{w,1}(\tilde{z})= n_{1,0}\, {{1}\over{\tilde{z}^2 + \tilde{r}^2}}\,,
\label{eq29}
\end{eqnarray}
\noindent
where $n_{1,0} = \rho_{1,0} / \bar{m}$ ($= \dot{m_1}/ 4 \pi \bar{m}
v_{1,0} R_0^2$ ) is the number density of the flow at the stagnation point.
From equations (\ref{eq25}), (\ref{eq28}), and (\ref{eq29}), it follows
that
\begin{eqnarray}
\tau_{w,1} (\theta)= \biggl({{\nu}\over{\nu_c}}\biggr)^{-2.1}
\,I_{w,1}(\theta)\,,
\label{eq30}
\end{eqnarray}
with
\begin{eqnarray}
I_{w,1}(\theta)= {{1}\over{2 \tilde{r}^3}}\,
\biggl[{{\pi}\over{2}} + {{\mbox{ctg}\,\theta}
\over{\mbox{ctg}^2\,\theta + 1}} + \theta\,\biggr]\,.
\nonumber
\end{eqnarray}

\noindent
Analogously, the optical depth of the wind source located at $z= D$ is
calculated by
\noindent
\begin{eqnarray}
\tau_{w,2} (\theta)=
 \int_{\tilde{r}\, \mbox{ctg}\, \theta}^{\infty} n_{w,2}^2(z)\,\chi(\nu)\,dz\,,
\label{eq31}
\end{eqnarray}
where
\begin{eqnarray}
n_{w,2}(\tilde{z})= n_{2,0}\, {{1}\over{[\tilde{z}-\tilde{D}]^2 +
\tilde{r}^2}}\,,
\label{eq32}
\end{eqnarray}
where $n_{2,0} =\rho_{2,0}/\bar{m}$ ($=\dot{m_2}/ 4 \pi \bar{m} v_{2,0} R_0^2$)
is the number density of the wind at the stagnation point. Substitution of
(\ref{eq32}) into equation (\ref{eq31}) gives
\begin{eqnarray}
\tau_{w,2} (\theta)= \biggl({{\nu}\over{\nu_c}}\biggr)^{-2.1}
\,I_{w,2}(\theta)\,,
\label{eq33}
\end{eqnarray}
\noindent
with
\begin{eqnarray}
I_{w,2}(\theta)= \biggl({{\dot{m_2}\,v_1}\over{\dot{m_1}\,v_2}}\biggr)^2
{{1}\over{2 \tilde{r}^3}}\,
\biggl[{{\pi}\over{2}} -
 {{(\tilde{r}\,\mbox{ctg}\,\theta - \tilde{D})\,\tilde{r}}
 \over{\tilde{r}^2 + (\tilde{r}\,\mbox{ctg}\,\theta - \tilde{D})^2}}
\nonumber
\end{eqnarray}
\begin{eqnarray}
 - \mbox{arctg}\,\biggl({{\tilde{r}\,\mbox{ctg}\,\theta - 
 \tilde{D}}\over{\tilde{r}}}\biggr)\,\biggr]\,.
\nonumber
\end{eqnarray}

We assumed that the system is far enough ($D \ll L$, being
$L$ the distance to the observer) so that
the lines of sight intersecting the central stars can be ignored. Then, the
radio-continuum flux from the colliding wind binary can be calculated by
\begin{eqnarray}
S_{\nu}= 2 \pi B_{\nu}\,\biggl({{R_0}\over{L}}\biggr)^2\,
\int_{0}^{\tilde{r}(\theta_{\infty})} [1 - \mbox{e}^{-\tau(\theta,\theta_1)}]\,
\tilde{r}\,d\tilde{r},
\label{eq34}
\end{eqnarray}
\noindent
where $\tau(\theta,\theta_1)= \tau_{WCR}(\theta, \theta_1) + \tau_{w,1} (\theta)
+ \tau_{w,2} (\theta)$ is the total optical depth along the line of sight,
$B_{\nu}\,(= 2 k T \nu^2/c^2$ with $k$ the Boltzmann's constant,
and $c$ the light speed) is the Planck function in the Rayleigh-Jeans
approximation, and $\tilde{r}(\theta_{\infty})$ is the impact parameter
at the asymptotic angle $\theta_{\infty}$ (corresponding to $R \rightarrow
\infty$) of the thin shell, which can be found from
\begin{eqnarray}
\theta_{\infty} - \mbox{tan}\,\theta_{\infty} = {{\pi}\over{1-\beta}}\,
\nonumber
\end{eqnarray}
\noindent
(see Cant\'o, Raga $\&$ Wilkin 1996). Defining the parameter $f_c =
4 \pi k T (R_0/ c L)^2 \nu_c^2$, we finally obtain 
\begin{eqnarray}
S_{\nu}= f_c \,\biggl({{\nu}\over{\nu_c}}\biggr)^{2}\biggl[
\int_{0}^{\theta_{\infty}} [1 - \mbox{e}^{-\tau(\theta,\theta_1)}]\,
\tilde{R}^2(\theta)\,\mbox{sin}\,\theta\,\mbox{cos}\,\theta\,d\theta\, +
\nonumber
\end{eqnarray}
\begin{eqnarray}
\int_{0}^{\theta_{\infty}} [1 - \mbox{e}^{-\tau(\theta)}]\,
\tilde{R}(\theta)\,\mbox{sin}^2\,\theta\,
(\partial \tilde{R}/ \partial \theta)\,d\theta\,\biggr]\,.
\label{eq35}
\end{eqnarray}
\noindent
Equation (\ref{eq35}) represents an self-similar solution for the
free-free radiation from colliding stellar winds from massive binary
systems.

   \begin{figure}[t!]
   \centering
   \includegraphics[width=9cm]{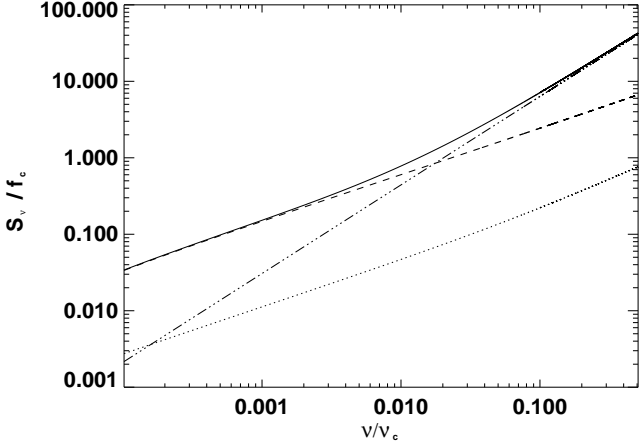}
      \caption{Predicted free-free emission from a colliding wind binary.
We have adopted the parameters $v_1= 10^{3}$ km s$^{-1}$,
$\dot m_{1} = 1.25 \times 10^{-5}$ M$_{\odot}$ yr$^{-1}$, and $v_2= 10^{3}$
km s$^{-1}$, $\dot m_{2} = 5 \times 10^{-5}$ M$_{\odot}$ yr$^{-1}$ for the wind
sources. The binary separation is set to $D=$ 4 AU. The dotted and
dashed lines represent the fluxes ($\propto \nu^{0.6}$) from the wind sources
1 and 2 (see also Fig. 1.), respectively. The intrinsic thermal emission from
the WCR (dot-dashed line) and the total emission (solid line) from the binary
system are also shown. The behavior of the curves is described in the main text.}
         \label{f3}
   \end{figure}

We apply the model for computing the emission from a
colliding wind binary using stellar wind parameters
similar to those from a typical WR+O binary system. That is, we have
adopted  $v_1= 10^{3}$ km s$^{-1}$, $\dot m_{1} = 1.25
\times 10^{-5}$ M$_{\odot}$ yr$^{-1}$ for the O type star, and for the WR component,
$v_2= 10^{3}$ km s$^{-1}$, $\dot m_{2} = 5 \times 10^{-5}$ M$_{\odot}$ yr$^{-1}$.
In addition, we have assumed a separation $D=$4~AU between the stars
(so that the WCR is radiative; see $\S$ 3).
The result is shown in Figure \ref{f3}.
It shows the contribution to the total radio emission
of the unshocked stellar winds and the WCR. As expected
for ionized stellar envelopes (see $\S$ 1), the flux densities from
the stellar winds increases as $\sim \nu^{0.6}$. However, deviation
from the expected 0.6 value of the spectral index at high frequencies
is observed from the emission of the wind source 1. This is probably
due to the presence of the WCR inside the optically thick region
of the wind. We note that as $\nu$ increases the emission from the WCR
becomes more important. The intrinsic flux density from the WCR shows a
spectral index of $\sim 1.1$, consistent with the numerical models
(in massive O+O type binary stars) developed by Pittard (2010). 
At low frequencies, the total flux density grows as $S_{\nu}\sim \nu^{0.6}$
approaching the emission from the stronger wind. On the other hand,
at higher frequencies, the radio spectrum from the system approaches
the flux density from the WCR ($S_{\nu}\sim \nu^{1.1}$).

   \begin{figure*}{t!}
   \centering
   \includegraphics[width=17.5cm]{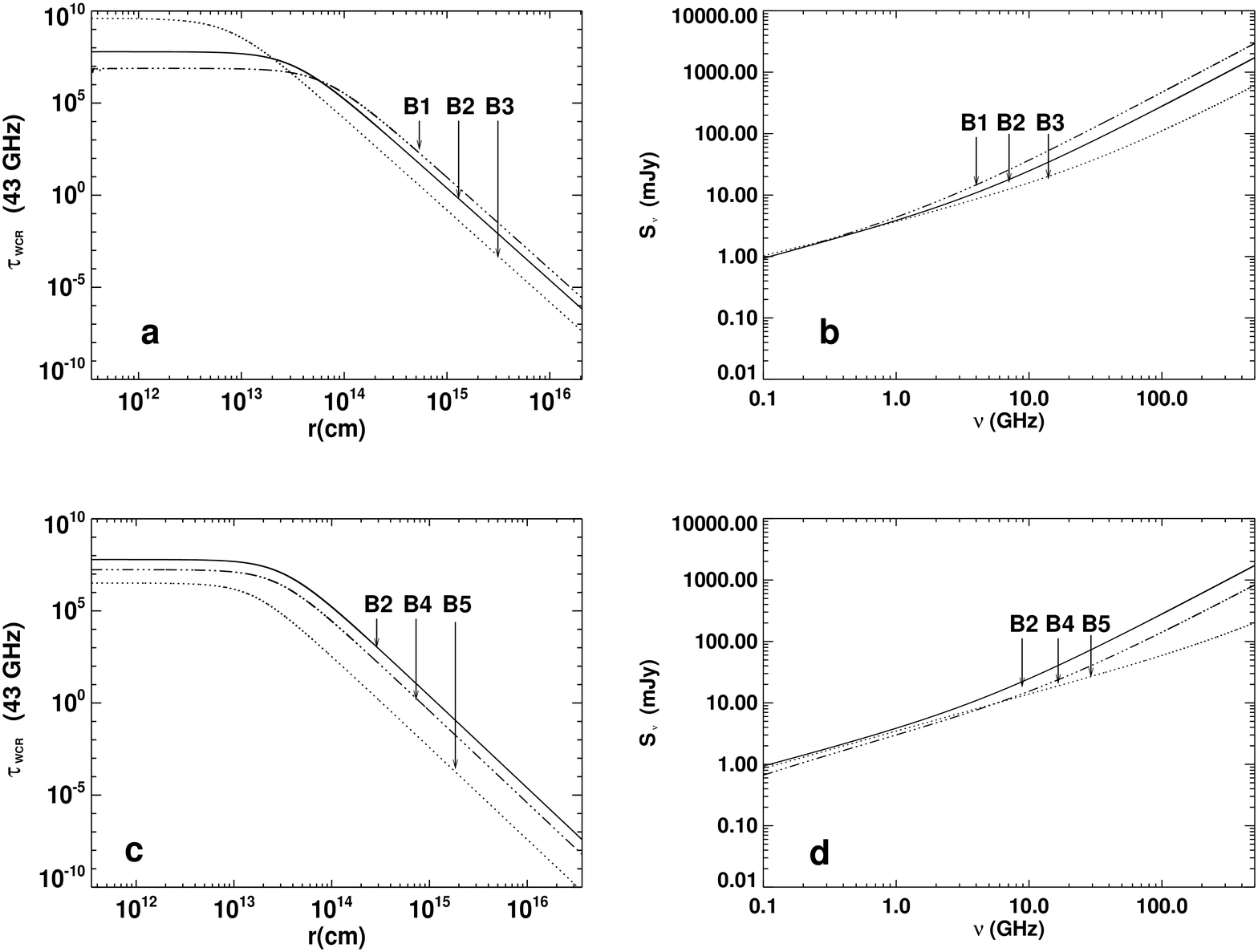}
      \caption{Optical depth $\tau_{WCR} (\mbox{43 GHz})$ of the WCR as a function 
of impact parameter $r$ (left panels), and the predicted flux density $S_{\nu}$
at radio frequencies (right panels) for the models of Table 1. In top panels,
the dot-dashed, solid and dotted lines represent the optical depth and the radio spectrum
from the models B1, B2, and B3, respectively. In bottom panels, the corresponding
values of models B2, B4, and B5 are shown by the solid, dot-dashed, and dotted lines,
respectively. The physical description of the plots are given in the text.
      }
         \label{f4}
   \end{figure*}

\section{Thermal radio-continuum emission from colliding wind binaries}

As mentioned in $\S$ 1, radio observations (e.g. Moran et al. 1989;
Dougherty, Williams $\&$ Pollaco 2000; Montes et al. 2009) and theoretical
models (see, for instance, Dougherty et al. 2003; Pittard et al. 2006;
Pittard 2000) of massive binary stars have revealed 
strong shocks formed in the wind-wind interaction zone. These shocks
 emit radio-continuum radiation consistent with thermal emission that
may produce variations in the expected flux densities and spectral
indices from ionized stellar winds. In this section, we apply the
model of colliding-wind binary systems developed in $\S$ 2 by adopting
different wind parameters of the components. Our model assumes strong
radiative shocks which collapse onto a thin shell, so that the width of
the WCR can be neglected. We investigate the contribution of the WCR to
the thermal radiation from colliding-wind binaries, and the effect of the
binary separation on the radio spectrum.

Stevens, Blondin $\&$ Pollock (1992) investigated the collision of
stellar winds in early-type binary systems. They studied the role of
radiative cooling in the structure and dynamics of colliding wind
binaries. In this work, the importance of cooling in
a particular system is quantified using the cooling parameter $\chi$
($\approx 0.15 \, v_3^4\,d_{\rm{AU}}/\dot{M}_5$, where $v_3$ the wind
velocity in units of $10^3\,\rm{km}\,s^{-1}$, $d_{\rm{AU}}$ the distance
to the contact discontinuity in units of AU, and $\dot{M}_5$ is the mass
loss rate in units of $10^{-5}\,\rm{M_\odot \, yr^{-1}}$), defined as the
ratio of the cooling time of the shocked gas to the escape time from
the intershock region. For models with $\chi \ge$ 1, the postshock
flow can be assumed to be adiabatic, while it is roughly isothermal for models with $\chi \ll$ 1. Numerical simulations by these authors show
that the cooling result in the formation of a thin dense shell, confined by
isothermal shocks, as $\chi$ drops below unity.
Antokhin, Owocki, $\&$ Brown (2004) have also investigated the narrowness of
the cooling layer using the ratio $l_0/ R$ (where $l_0$ is the cooling length
and $R$ the radius from the star source), which is closely related to the
above parameter $\chi$. These authors show that the ratio $l_0/ R$
can serve as well to distinguish between adiabatic and radiative shocks,
with $l_0/ R>1$ implying an adiabatic shock and $l_0/ R < 1$ a radiative
one.

Nonetheless, radiative pressure that moderates the wind-wind
collision may be important in close hot-star binaries. An initial analysis
was developed by Stevens $\&$ Pollock (1994), who investigated the dynamics
of colliding winds in massive binary systems. They show that, in
close binaries, the radiation of a luminous star inhibits
the initial acceleration of the companion's wind towards the stagnation
point. These radiative forces result in lower velocities than those
expected in single-star models moderating the wind collision. 
In other work, Gayley, Owocki $\&$ Cranmer (1997) investigated the potential
role of the radiative braking effect, whereby the primary wind is decelerated
by radiation pressure as it approaches the surface of the companion star.
These authors conclud that radiative braking must have a significant
effect for wind-wind collision in WR+O binaries with medium separations
($D<$ 0.5 AU). Furthermore, Parkin and Pittard (2008) carried out
3D hydrodynamical simulations of colliding winds in binary systems.
They shows that the shape of the WCR is deformed by Coriolis forces into
spiral structures by the motion of the stars. The shape of the shock layer
is more deformed in systems with eccentric orbits.
In addition, Pittard (2010) developed 3D hydrodynamical models of
massive O+O star binaries for computing the thermal radio to
submillimeter emission. In this work, flux and spectral index
variations with orbital phase and orientation of the observer
(from radiative and adiabatic systems) are investigated. 
These authors found strong variations in eccentric systems
caused by dramatic changes to the physical properties (density
and temperature) of the WCR, which is radiative (optically thick)
at periastron, but adiabatic (optically thin) at apastron.

\subsection{The predicted radio emission from radiative shocks in
 colliding wind binaries}

Here, we present analytic predictions of the thermal spectra
at radio frequencies from different radiative models, which
satisfy the condition of radiative shocks ($\chi<$ 1)
for the WCR. 
We have estimated the value of $\chi$ for each shock of the WCR.
In Table 1, we list the different scenarios of colliding
wind binaries that we have studied in this paper.
In models B1-B3, we have assumed identical wind sources ($\beta$= 1)
in order to investigate the effect of the binary separation. On the other
hand, in models B2, B4, and B5, we have assumed different wind momentum
ratios for the binary systems, while the distance between the components
is unchanged. In models B4 and B5, we indicate the highest value of $\chi$ for each system.
These models for the radio emission from colliding
wind binaries are presented in Figure \ref{f4}. We show the optical
depth of the WCR as functions of the impact parameter $r$ (left panels)
and the predicted flux density at radio frequencies (right panels).
Top panels show the results of our Models B1-B3, while the models
B2, B4, and B5 are shown in the bottom panels. All models are calculated
for an observer located in the orbital plane along the symmetry axis
(which corresponds to a system with an inclination angle $i=90^{\circ}$
with the most powerful stellar wind in front, as shown in Figure 1).

\begin{table}
\begin{center}
TABLE 1
\centerline{\sc Parameters of the colliding winds}
\centering                          
\begin{tabular}{c c c c c}        
\hline
\hline 
Model $\, ^{(a)}$& $\beta$ &$D$/AU & $\chi$ & $R_0/D$ \\  
\hline 
   B1 & 1 & 8 & 0.12    & 0.5  \\
   B2 & 1 & 4 & 0.06    & 0.5  \\
   B3 & 1 & 1 & 0.02    & 0.5  \\ 
\hline 
Model $\, ^{(b)}$& $\beta$ &$D$/AU & $\chi$ & $R_0/D$ \\  
\hline 
   B4 & 0.25 & 4 & 0.15  & 0.33 \\
   B5 & 0.025 & 4 & 0.65  & 0.14 \\
\hline\hline 
\end{tabular}
\begin{flushleft}
{(a) $v_{1,2} = 10^{3}$ km s$^{-1}$; $\dot m_{1,2} = 5 \times 10^{-5}$ M$_{\odot}$ yr$^{-1}$}\\
{(b) $v_{1,2} = 10^{3}$ km s$^{-1}$; $\dot m_{2} = 5 \times 10^{-5}$ M$_{\odot}$ yr$^{-1}$; $\dot m_{1} = 1.25 \times 10^{-5}$ M$_{\odot}$ yr$^{-1}$ for model B4 and  $\dot m_{1} = 1.25 \times 10^{-6}$ M$_{\odot}$ yr$^{-1}$ for model B5.} \\
\end{flushleft}
\end{center}
\end{table}
%

First, we observe that, for low-impact parameters ($r/R_0 \leq 1$), the
optical depth $\tau_{WCR} (\mbox{43 GHz})$ of the WCR (Figure \ref{f4}-a)
does not depend on the value of $r$ and scales as $D^{-3}$. On the other hand,
for higher impact parameters ($r/R_0 \gg 1$), $\tau_{WCR} (\mbox{43 GHz})
\propto r^{-5}$ in all models. At a given line of sight, we also note that,
in this limit, the optical depth increases as $D^{2}$. The transition impact
parameter is $r\propto R_0$ (see, also, Appendix A).
This behavior of the optical depth (and therefore of the emission
measure of the thin shell) can be explained as follows (see Figs. \ref{f7}
and \ref{f8}). We consider a given impact parameter $r$ and vary the binary
separation. For systems with $r/R_0 \gg 1$ ($\tau_s \propto r^{-5}$), as
the binary separation (and therefore $R_0$) increases, the ratio $r/R_0$
decreases. The normal components of the shock velocities (for the two shock
fronts that bound the WCR) increase for higher values of $R_0$, which
result in higher pressures within the shocked layer (eqs. [\ref{eq8}] and
[\ref{eq9}]). Thus, both the emission measure (eq. [\ref{eq15}]) and the
optical depth (eq. [\ref{eq27}]) of the thin shell increase.
On the other hand, for systems with $r/R_0 \ll 1$ ($\tau_s \neq
\tau_s [r/R_0]$), as $R_0$ increases the preshock densities (for the two
shock fronts of the WCR) decreases, which result in lower pressures
within the shocked layer. Consequently, the emission measure and the
optical depth of the WCR diminish.

Second, we show the  predicted flux density at radio frequencies
for models B1-B3 (Figure \ref{f4}-b). We note that, at frequencies
$\nu >$ 1 GHz, as the binary separation increases, the emission from the
binary becomes more important. In order to explain this behavior of
the spectrum, we investigated the emission from the shocked layer
as a function of the impact parameter $r$ (see $\S$A.3 of the Appendix A).
Our results are presented in Fig. \ref{f9}. We show that the main
contribution to the flux comes from the optically thick region of the
shell, so the emission from the optically thin region can be
neglected. In addition, our model predicts that the emission from an
optically thick WCR (which is bounded by radiative shocks) scales as
$S_\nu \propto D^{4/5}$ (see Fig. \ref{f10}). This result contrasts
with the inverse dependence with the binary separation described by
Pittard et al. (2006) for an adiabatic and optically thin WCR, where
the emission is expected to scale as $D^{-1}$.

Finally, as mentioned above, the effect of varying the wind momentum
ratio $\beta$ on the value of $\tau_{WCR}(\mbox{43 GHz})$, and the total
spectrum, is investigated from models B2, B4, and B5 (bottom panels of
Fig. \ref{f4}). As we expected from equations (\ref{eq26})-(\ref{eq27}), it can
be observed that the optical depth of the WCR increases with the mass loss
rate $\dot m_1$ of the secondary component (Fig. \ref{f4}-c). In addition,
Figure \ref{f4}-d shows the predicted thermal radio spectra from these models.
The plots show how sensitive our analytic model is to the wind momentum
radio, $\beta$. We note that, at high frequencies, the total flux density
$S_{\nu}$ increases with the parameter $\beta$.

\begin{table}
\begin{center}
TABLE 2
\centerline{\sc Wind parameters of colliding wind models}
\centering                          
\begin{tabular}{c c c c c c}        
\hline   
\hline               
Model     & $\beta$ & D$/AU$ & $\chi$ & $R_0/D$ & $\theta_\infty/\pi$\\    
\hline                        
   WR 98  & 0.20    & 0.5     & 0.6    & 0.31  & 0.6  \\      
   WR 113 & 0.21    & 0.5     & 0.6    & 0.31  & 0.6   \\
\hline\hline 
\end{tabular}
\end{center}
\end{table}

\section{Comparison with observations}

As we mention in previous sections, the model presented in this
paper assumes a thin shell approximation for the colliding wind region.
In the presence of efficient cooling ($\chi <1$), the WCR is confined
by radiative shocks, which are expected in close binaries. Consequently,
we investigate the emission from short period ($<$ 1 yr) WR+O systems
calculated with the analytic model developed in $\S$ 2.
Besides, radiative breaking can be important for binaries with a
medium separation  $D  <100\, R_\odot$ (orbital periods $\sim$15 days),
which prevents the applicability of our model for systems with shorter
periods.

 Recently, Montes et al. (2009) have reported
observations of the flux density and spectral indices for the close systems
WR~98 113, 138, and 141. In this section, we apply the analytical model to
the binaries WR 98 and WR 113 and compare them with observations. Table 2 lists
the adopted stellar parameters of these sources in our colliding wind model,
which suggest the presence of radiative shocks in the wind-wind interaction
zone. We indicate the highest value of $\chi$ for each binary system.
We indicate the highest value of $\chi$ for each binary system.
We show that the contribution of the WCR to the total emission gives a
possible explanation to the observations.

\textit{WR~98}.
This close binary has been identified as a double line binary (WN7o/WC+O8-9) with an
orbital period, $P=47.8$~days (Gamen \& Niemela 2002). Abbott et al. (1986) classified
WR~98 as a nonthermal radio source. 
 Recently, multi-wavelength observations (from 5 to 23~GHz) of this source 
have revealed  that the spectral index changed from $\sim 0.26$ to $\sim 0.64$ in a period of time of $\sim 15$~days (Montes et al. 2009).  This behavior was interpreted as indicating a binary influence over the radio spectrum, possibly from a variable nonthermal contribution that is absorbed at certain orbital phases, turning the spectrum into a ``thermal state''.
The stellar parameters of this binary indicate that a radiative WCR ($\chi \sim 0.6$) is formed within this system.

   \begin{figure}[t!]
   \centering
   \includegraphics[width=9cm]{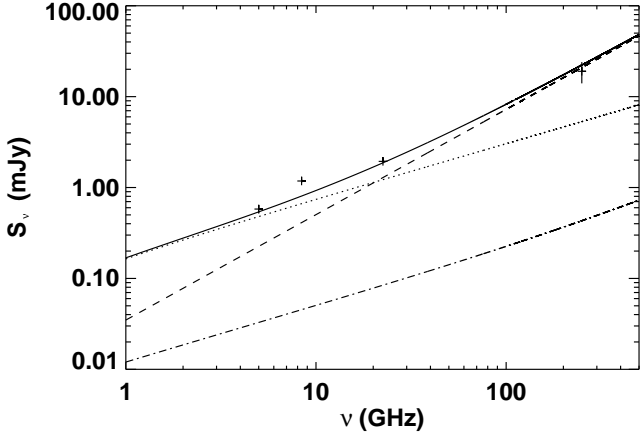}
      \caption{Comparison between our model and the flux densities at
5, 8.4, 23 and 250 GHz ('+' sign; uncertainty in the flux measurements are given by the length of the bars) of the binary
system WR~98 obtained by Montes et al. (2009) and Altenhoff et al. (1991),
respectively.  We have assumed the stellar wind parameters $\dot{M}_{WR}=3.0
\times 10^{-5}\,\rm{M_\odot\, yr^{-1}}$, $v_{WR}=1200\,\rm{km\,s^{-1}}$,
$\dot{M}_{O}=4.0 \times10^{-6}\,\rm{M_\odot\, yr^{-1}}$, and $v_{O}=1800\,
\rm{km\,s^{-1}}$. The stars are separated by a distance $D\sim 0.5$~AU.
The dotted and dashed lines represent the fluxes ($\propto \nu^{0.6}$) from
the winds of the WR and O type stars, respectively. The radiation from the WCR
(dot-dashed line) and the total flux density (solid line) from the binary
source are also shown. The physical description of the plot is given in the
text.}
         \label{f5}
   \end{figure}

We applied the model to the binary system WR~98 to investigate if a thermal component of emission  from the WCR  is able to contribute significantly to the total spectrum. Our results from the model are compared with the observations  at the thermal state ($\alpha \sim 0.64$) reported by Montes et al. (2009).
We assumed for the WR star a mass loss rate $\dot{M}_{WR}=3.0 \times 10^{-5}\,
\rm{M_\odot\, yr^{-1}}$ (upper limit derived from radio observations at the thermal phase
at 8.4~GHz when $\alpha \sim 0.64$; Montes et al. 2009), and an ejection velocity
$v_{WR}=1200\,\rm{km\,s^{-1}}$ (Eenens \& Williams, 1994). 
The best fit to the observations was found from the O star parameters $\dot{M}_{O}=4.0 \times10^{-6}\,\rm{M_\odot\, yr^{-1}}$
and $v_{O}=1800\,\rm{km\,s^{-1}}$.
 We have assumed a mean atomic weight per
electron $\mu=4.2$ and an average ionic charge $Z=1.1$ for the WR wind, and $\mu=1.5$ and $Z=1$ for the O star (which are typical values for O-type stars; see Bieging et al. 1989). 
Using equation (19) in Cant\'o et al. (1996) and considering that most of the emission from the WCR arise from impact parameters, $r$, such that $\theta < \theta_\infty$ (see Table 2), we can estimate the ratio of mass entering into the WCR from the WR stellar wind, to that from the O star, $\Delta M_{WR}/\Delta M_O > 0.75$.
Thus, we assume that the material within the shock is mainly composed of what comes from the strongest wind, assuming the same values of $\mu$ and $Z$ for the shocked material as those used for the WR wind.
   From the value of $a \sin{i} \sim 100 \, R_\odot$, (where $a$ is the semi-major axis and $i$ is the inclination angle of the orbit)  determined by Gamen \& Niemela (2002), we assume $D\sim 0.5$~AU for the binary separation, which is the maximum value for $a$ (when $i\sim 90^\circ$).  Since this value represents an upper limit for $D$ and because the geometric configuration represents the WR stellar wind in front of the O star, we are determining the minimum expected for the WCR thermal contribution.

In Figure \ref{f5}, we show the predicted spectrum from WR~98 system. Observational
data ($S_{5\,\rm{GHz}}=0.58 \pm 0.06 \,$mJy, $S_{8.4\,\rm{GHz}}=1.18 \pm 0.05 \,$mJy,
and $S_{23\,\rm{GHz}}=1.94 \pm 0.15 \,$mJy) by Montes et al. (2009) are also plotted. 
It can be seen that our model predicts an increase in the flux density at high frequencies, which results in
steeper spectral indices than the expected 0.6 value of the stellar winds.
Also the 
 WCR emission becomes comparable to that from the WR wind at
$\sim 70$ GHz, and at a frequency of 250 GHz, the shock layer produces an excess
of emission of a factor $>2$ over the value expected for the WR wind (8~mJy). 
To test this  prediction  of the model at higher frequencies,
we included the flux density $S_{250\,\rm{GHz}}=19 \pm 5\,$mJy
measured by Altenhoff et al. (1991), which seems to agree with the 22~mJy predicted for the total flux density at 250~GHz. Furthermore, the spectral index of the WCR emission , $\alpha_{\rm{WCR}}\sim 1.2$, is similar to what is derived from the observations at this frequency range, $\alpha_{22.5-250\,\rm{GHz}}\sim 0.95$.

   \begin{figure}[t]
   \centering
   \includegraphics[width=9cm]{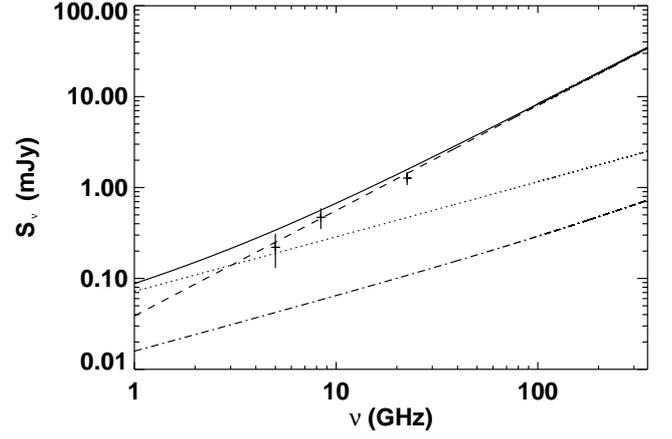}
      \caption{Comparison between our model and the flux densities at
5, 8.4, 23 GHz ('+' sign; uncertainty in the flux measurements are given by the length of the bars) of the binary system WR~113 obtained by Montes
et al. (2009). We have assumed the stellar wind parameters $\dot{M}_{WR}=
2.0\,\times 10^{-5}\,\rm{M_\odot\, yr^{-1}}$, $v_{WR}=1700\,\rm{km\,s^{-1}}$,
$\dot{M}_{O}=0.4\times 10^{-5}\rm{M_\odot\, yr^{-1}}$, and $v_{O}=1800\,
\rm{km\,s^{-1}}$. The stars are separated by a distance $D\sim 0.6$~AU.
The dotted and dashed lines represent the fluxes ($\propto \nu^{0.6}$) from
the winds of the WR and O type stars, respectively. The radiation from the WCR
(dot-dashed line) and the total flux density (solid line) from the binary
source are also shown. The physical description of the plot is given in the
text.
      }
         \label{f6}
   \end{figure}


\textit{WR~113}. This source was detected at radio frequencies (8.4~GHz) for first
time by Cappa et al. (2004). The period of this spectroscopic binary (WC8+O8-9) is $P=29.7$~days (Niemela et
al. 1996). Recently, simultaneous multifrequency observations (from 5 to 23~GHz)
of WR~113 by Montes et al. (2009) showed an spectral index $\alpha\sim 1.06$.
We have modeled this binary adopting the wind parameters $\dot{M}_{WR}=2.0
\,\times 10^{-5}\,\rm{M_\odot\, yr^{-1}}$ (Lamontagne et al. 1996) and $v_{WR}=1700\,\rm{km\,s^{-1}}$ (van der Hucht et al. 2001) for the WR star, and $\dot{M}_O=0.4\times 10^{-5}\rm{M_\odot\, yr^{-1}}$ and $v_0=1800\,
\rm{km\,s^{-1}}$ for the O type component, separated by a distance $D=0.6$~AU
(from the value $a\sin{i} = 129\, R_\odot$, and $i \sim 70$ determined by Lamontagne et al. 1996). We assume a mean atomic weight per electron
$\mu=4.7$ and an average ionic charge $Z=1.1$ for the stellar wind of the WR wind (from Cappa et al. 2004), and $\mu=1.2$ and $Z=1$ for the O star. As in the case of WR~98, we assumed that the shock is mainly composed of material from the WR wind ($\Delta M_{WR}/ \Delta M_O > 0.65$) , and the same $\mu$ and $Z$ values were used.
Figure \ref{f6} shows the predicted spectrum by the model, as well as observational data
($S_{5\,\rm{GHz}}= 0.22 \pm 0.03 \,$mJy, $S_{8.4\,\rm{GHz}}= 0.47 \pm 0.04\,$mJy,
and $S_{23\,\rm{GHz}}=1.27 \pm 0.07 \,$mJy) from the WR~113 system
by Montes et al. (2009). 
We note from the figure that the contribution of the WCR
starts to dominate the emission from the system at a lower frequency ($\nu\sim 9$~GHz) than in the case of WR~98, mainly because the flux density from the WR star is $\sim3$ times lower than for WR~98.
At a  frequency of 250 GHz, the flux density from the thin shell becomes a factor
of $\sim 6$ greater that the expected value from the unshocked winds. Observations
at such high frequencies are required in order to verify the high spectral indices
predicted by our model.

From our model, we have shown that a thermal contribution from the WCR in WR~98 and WR~113 is likely to be detected at high frequencies. However, as we pointed out, we can only present a lower limit for this contribution, owing to the uncertainty of parameters such as the binary separation, $D$.
On the other hand, Pittard (2010) have investigated the variability
in the flux density (from radiative systems) due to the orbital motion of
the stars and orientation of the observer. The configuration adopted
in our model (with an inclination angle of $90^\circ$ and orbital phase
$\sim$ 0) corresponds to the minimum contribution to the total thermal
emission expected from radiative shocks in binary systems  with an inclination angle $\sim 90^\circ$.
In this way, the excess of emission predicted here is expected to be variable and modulated by the orbit motion. 
WR~98 was reported as a variable source by Montes et al. 2009, changing its spectral index from $\sim$0.26 to $\sim$0.64 in a period of $\sim$15~days. Although the flat spectral index ($\sim$0.26) resembles the one predicted by Pittard et al. 2006 for an adiabatic shock, this close system is likely to be radiative, and such behavior was explained as result of a nonthermal contribution escaping the absorption.
On the other hand, for WR~113 there are no observations at high frequencies or clear evidence of variability to support the contribution predicted here; therefore, high-frequency observations are required to confirm it.

\section{Summary and conclusions}
In this paper, we have presented an analytical model for calculating the thermal contribution at radio frequencies from the WCR in massive binary systems. The strong
stellar winds of both components collide and form the WCR (bounded by two
shock fronts) between the stars, which emits both thermal (free-free) and
nonthermal (synchrotron) radiation. In particular, the nonthermal
component is expected to be highly absorbed in close binaries, so
the thermal flux must be the dominant contributor. In addition, particle acceleration should not
occur if the shocks are collisional such as in short period systems
where the gas density is high.

Based on considerations of linear and angular momentum conservation, we
developed a formulation for calculating the emission measure of the WCR,
from which simple predictions of the thermal emission can be made.
We assumed that the WCR is bounded by radiative shocks, so that
the width of the WCR can be neglected and the wind-wind interaction
zone can be approximated as a thin shell. Since highly radiative shocks
are expected to occur in short period systems, we assumed different
scenarios of colliding wind close binaries. Using typical parameters of
massive WR+O binaries, we showed that the compressed layer emits free-free
radiation that plays a very substantial role in the emission from the
whole system. The radio-continuum spectrum obtained by our model clearly
deviates from the behavior $S_{\nu} \propto \nu^{0.6}$ predicted by the
standard stellar wind model.

In previous models (Pittard et al. 2006), the expected thermal emission
from binary systems where the WCR remains optically thin (confined by
adiabatic shocks) was investigated. In this case, the flux density scales
as $D^{-1}$; consequently, the emission from the WCR increases as the
binary separation decreases. In this work, we study the thermal radiation
from systems where the WCR is highly radiative. The effects of the binary
separation and the wind momentum ratio on the total spectrum are investigated.
Our model's results show that the relative contribution of the WCR to the
total emission mainly arises from the optically thick region of the layer,
which flux density scales as $D^2$ (where $D$ is the stellar separation).
Thus, the impact of the WCR to the total spectrum (in contrast to the
optically thin case) becomes more important as the binary separation
increases (until the WCR becomes optically thin).

Finally, we applied the analytical model to 
massive binaries, and calculated the thermal radio-continuum emission from the short-period 
WR+O systems, WR 98 and WR 113. Assuming a set of wind parameters
consistent with observations of these sources, we find that the WCR
must be confined by radiative shocks, and so the thin shell approximation
can be applied. In both sources, the compressed layer generates thermal
radiation that produces (at high frequencies) an excess of emission over
the expected values from the stellar winds. Comparison with recent
observations (by Montes et al. 2009) from these sources showed that our
model can satisfactorily reproduce the flux densities and spectral
indices.

\begin{acknowledgements}
G.M. acknowledges financial support from a CSIC JAE-PREDOC fellowship. RFG
has been partially supported by DGAPA (UNAM) grant IN117708.  JC acknowledges
support from CONACyT grant 61547. This research was partially supported by the
Spanish MICINN through grant AYA2009-13036-CO2-01. The authors acknowledge
the anonymous referee for the helpful comments that improved the content
and presentation of the paper.     
\end{acknowledgements}

\begin{appendix}

\section{Emission from a binary system with identical stellar winds}

We consider two stars with identical winds. The stellar winds
are ionized and isotropically ejected with terminal velocity
$v\, (= v_1= v_2)$ and mass loss rate $\dot m\, (\dot m_1= \dot m_2)$.
In this particular case, the wind momentum ratio of the interacting
winds $\beta= 1$, the stagnation point radius $R_0= D/2$, and the
angle $\theta= \theta_1$ (see Fig. 1).

First, we estimate the emission measure of the shocked layer
as follows. From equations (\ref{eq18}), (\ref{eq20}), and
(\ref{eq21}), one obtains
\begin{eqnarray}
f_m(\theta)= 2\,(1 - \mbox{cos}\,\theta)\,,
\label{a1}
\end{eqnarray}

\begin{eqnarray}
f_r(\theta)= \theta -  \mbox{sin}\,\theta\, \mbox{cos}\,\theta\,,
\label{a2}
\end{eqnarray}
\noindent
and,
\begin{eqnarray}
f_z(\theta)= 0\, .
\label{a3}
\end{eqnarray}
Substitution of eqs. (\ref{a1})-(\ref{a3}) into eq. (\ref{eq19}) gives
the velocity along the shell,
\begin{eqnarray}
v= {{v_1}\over{2}}\, \biggl({{\theta -  \mbox{sin}\,\theta\,
 \mbox{cos}\,\theta}
\over{1 - \mbox{cos}\,\theta}}\,\biggr)\,.
\label{a4}
\end{eqnarray}
On the other hand, the surface density of the layer is calculated
as follows. First, we obtain the mass injection rate $\Delta \dot{M}$
(into a solid angle defined by an impact parameter $r$) of the stellar
winds,
\begin{eqnarray}
\Delta \dot{M}= \dot{m_1}\,(1 - \mbox{cos}\,\theta)\,.
\label{a5}
\end{eqnarray}

   \begin{figure}[t]
   \centering
   \includegraphics[width=8cm]{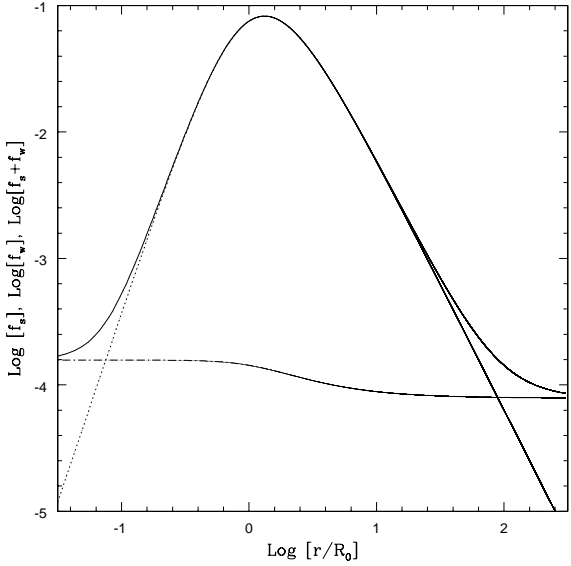}
      \caption{Nondimension emission measure as function of $r/R_0$.
The dotted and dashed lines represent the functions $f_s$ of the shocked
layer, and $f_w$ of the stellar winds, respectively. The total emission
measure $f_s + f_w$ is also shown (solid line). The physical description
of the plot is given in the text.}
         \label{f7}
   \end{figure}

Next, we combine equations (\ref{eq16}) and (\ref{a5}) in order to
find the surface density,
\begin{eqnarray}
\sigma= {{\dot{m_1}\,(1 - \mbox{cos}\,\theta)^2} 
\over {\pi\, v_1\, R \,\mbox{sin}\,\theta\,
(\theta -  \mbox{sin}\,\theta\,\mbox{cos}\,\theta)}}\,,
\label{a6}
\end{eqnarray}
\noindent
where we have used the velocity $v$ given by equation (\ref{a4}).

In this simple case $R= R_0 / \mbox{cos}\,\theta$, and the normal
component of the preshock velocity of the stellar wind $v_{1,n}=
v_1\, \mbox{cos}\,\theta$ (see eq. [\ref{eq6}]). Therefore,
the pressure $P_1\,= \rho_1\, v^2_{1,n}$ within the thin shell just
behind the shock front can be written as
\begin{eqnarray}
P_1= {{\dot{m_1}\,v_1}\over{4\pi\,R^2_0}}\,\mbox{cos}^4\theta\,.
\label{a7}
\end{eqnarray}
\noindent
Note from equations (\ref{eq7}) and (\ref{eq9}) that $P_1= P_2$,
as expected.
From equation (\ref{eq15}), it follows that the emission measure
of the shell is $EM_s= \sigma\,P_1/ \bar{m}^2\,c^2_s$. Using
equations (\ref{a6}) and (\ref{a7}), it can be shown that,
\begin{eqnarray}
EM_s= {{ \dot m^2_1 } \over{4\pi^2 \bar{m}^2 c^2_s R^3_0}}
 \biggl[{{\mbox{cos}^5\theta\,(1 - \mbox{cos}\,\theta)^2}
  \over{\mbox{sin}\,\theta\,(\theta -  \mbox{sin}\,\theta\,
 \mbox{cos}\,\theta)}}\biggr]\,.
\label{a8}
\end{eqnarray}

   \begin{figure}[t]
   \centering
   \includegraphics[width=8cm]{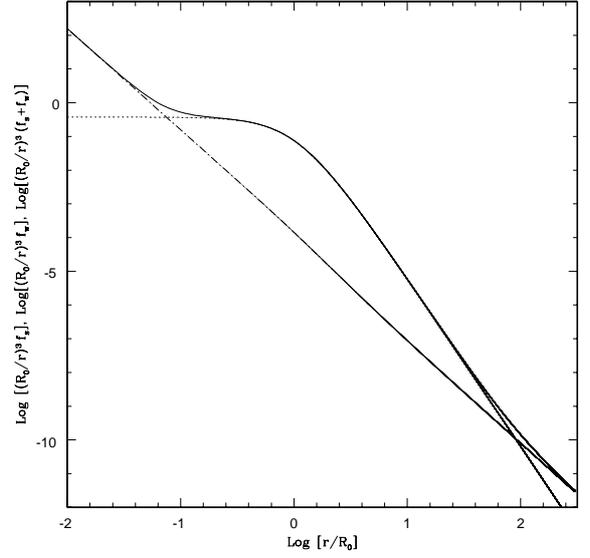}
      \caption{Nondimension emission measure as function of $r/R_0$.
The function $(r/R_0)^{-3}\,f_s$ (dotted line) of the shocked layer,
and $(r/R_0)^{-3}\,f_w$ (dashed line) of the stellar winds are presented.
The solid line represents the total emission measure $(r/R_0)^{-3}\,
(f_s + f_w)$. The physical description of the plot is given in the text.}
         \label{f8}
   \end{figure}

Let us now calculate the emission measure of the stellar
winds. Since we assume the same flow parameters of both
components of the binary system, then the emission measure
of the winds is given by
\begin{eqnarray}
EM_w= 2\, \int_{-\infty}^{R_0} n^2_{w} dz\,
 = 2\, \biggl({{\dot{m_1}}\over{4 \pi\, \bar{m} v_{1}}}\biggr)^2
 \,\int_{-\infty}^{R_0} {{dz}\over {(r^2 + z^2)^2}}\,,
\label{a9}
\end{eqnarray}
\noindent
and integrating one obtains
\begin{eqnarray} 
EM_w= {{\dot m^2_1}\over{16 \pi^2\,\bar{m}^2 v^2_1 r^3}}
\biggl [{{\pi}\over{2}} + {{(r/R_0)}\over{1 + (r/R_0)^2}} + 
\mbox{arctg}\biggl({{R_0}\over{r}}\biggr)
\biggr]\,.
\label{a10}
\end{eqnarray}
Thus, the total emission measure is given by $EM = EM_s+ EM_w$.

\subsection{The emission measure $EM$ as function of the stagnation
point radius $R_0$}

Let us now fix the impact parameter $r$, and investigate the behavior
of the emission measure $EM$ as function of the stagnation point radius
$R_0$. The total emission measure along a given line of sight can be
written as
\begin{eqnarray}
EM= {{\dot m_1^2}\over{4 \pi^2\,\bar{m}^2 c_s^2 r^3}},
\biggl (f_s + f_w \biggr )\,,
\label{a11}
\end{eqnarray}
\noindent
where
\begin{eqnarray}
f_s= \biggl ({{r}\over{R_0}}\biggr)^3 \,
  \biggl[{{\mbox{cos}^5\theta\,(1 - \mbox{cos}\,\theta)^2}
  \over{\mbox{sin}\,\theta\,(\theta -  \mbox{sin}\,\theta\,
 \mbox{cos}\,\theta)}}\biggr]\,,
\label{a12}
\end{eqnarray}
\noindent
and
\begin{eqnarray}
f_w= {{1}\over{4}}\, \biggl ({{c_s}\over{v_1}}\biggr)^2 \,
\biggl [{{\pi}\over{2}} + {{r/R_0}\over{1 + (r/R_0)^2}} + 
\mbox{arctg}\biggl({{R_0}\over{r}}\biggr)
\biggr]\,.
\label{a13}
\end{eqnarray}

It is also possible to give the emission measure $EM_s$ (eq.[\ref{a8}])
in terms of $r$ and $R_0$ by substituting the trigonometric functions
$\mbox{sin}\,\theta= (r/R_0)/\sqrt{1+(r/R_0)^2}$, $\mbox{cos}\,\theta=
1/ \sqrt{1+(r/R_0)^2}$, and $\theta= \mbox{arctg}(r/R_0)$ into equation
(\ref{a12}).

In Figure \ref{f7}, we present $f_s$ and $f_w$ as functions of $r/R_0$. We
fix the impact parameter $r$ and vary the stagnation point radius $R_0$. We
note from the figure that, initially as $r/R_0$ increases (which means closer
binary systems), the emission measure from the shell increases.
While $r/R_0$ is still growing, the emission measure reaches a maximum value
($f_s\simeq$ 1 at $r/R_0 \simeq$ 1.3) and then it steadily decreases.
In addition, it can be observed from the figure that $f_s \gg f_w$ at
$r/R_0\simeq 1$, and $f_s \ll f_w$ at very large ($r/R_0\ll 1$) or very
small ($r/R_0\gg 1$) stagnation point radii.

\subsection{The emission measure $EM$ as function of the impact
parameter $r$}

We now fix the stagnation point radius $R_0$ and investigate the variation
of the total emission measure with the impact parameter $r$. In this case,
it is useful to write the total emission measure $EM\, (= EM_s + EM_w)$ as 
\begin{eqnarray}
EM=\frac{\dot{m}_1^2 }{4 \pi^2 \bar{m}^2\, c^2 R_0^3} \,
 \biggl( \frac{R_0}{r} \biggl)^3 \, \biggl(f_s + f_w \biggl).
\label{a14}
\end{eqnarray}

  \begin{figure}[t]
   \centering
   \includegraphics[width=8cm]{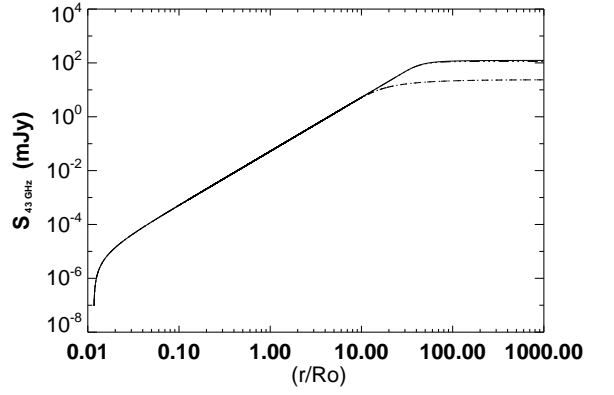}
      \caption{Integrated flux density at 43 GHz of the shocked
layer over the nondimensional impact parameter $r/R_0$. In this example,
we have assumed the parameters of the colliding winds of Model B2
(see Table 2). The behavior of the plot is described in the text.}
         \label{f9}
   \end{figure}

In Figure \ref{f8}, we show the contribution of the shell $EM_s$,
the contribution of the winds $EM_w$, and the total emission measure
$EM$ as functions of the nondimensional impact parameter $(r/R_0)$.
We note from the figure that the emission measure is dominated by
the shell ($EM_s\gg EM_w$) at impact parameters $(r/R_0)\simeq 1$.
However, at very low ($r/R_0\ll 1$) or very high ($r/R_0\gg 1$)
values of the impact parameter, the contribution of the stellar
winds to the total emission measure is more important ($EM_s\ll EM_w$). 

Given the emission measure $EM_s$ of the shocked layer, the optical
depth $\tau_{s}$ can be obtained by
\begin{eqnarray}
\tau_{s}= {{A_{\nu}}\over {r^3}}\,f_s
\label{a15}
\end{eqnarray}
\noindent
where $A_{\nu}= \dot m^2_1\,\chi(\nu)/(4\pi^2 \bar{m}^2 c^2_s)$.

For lines of sight with impact parameters $r/R_0\ll 1$, it can be
shown that $f_s\simeq 3/8\, (r/R_0)^3$, and
\begin{eqnarray}
\tau_{s}\simeq {{3}\over{8}}\,{{A_{\nu}}\over {R_0^3}}.
\label{a16}
\end{eqnarray}
\noindent
Consequently, with in this limit the optical depth of the shell $\tau_s$
does not depend on the value of $r$ and scales as $D^{-3}$ (being $D$
the binary separation). 

On the other hand, for higher impact parameters ($r/R_0\gg 1$)
$f_s\simeq 2/\pi \,(R_0/r)^2$ and
\begin{eqnarray}
\tau_{s}\simeq {{2}\over{\pi}}\,A_{\nu}{{R_0^2}\over{r^5}}.
\label{a17}
\end{eqnarray}
\noindent
Thus, in this limit ($\tau_s \propto r^{-5}$), the optical depth
of the shell (for a given line of sight) scales as $\tau_s \propto D^2$. 

\subsection{Flux density $S_{\nu}$ as function of the binary
 separation $D$}

In this section, we investigate the dependence of the flux density
 $S_{\nu}$ with the binary separation $D$. Figure \ref{f9} shows the
flux density $S_{43\,\rm{GHz}}$ of the shocked layer as function of the
nondimensional impact parameter $r/R_0$. We have adopted the wind
parameters of Model B2 (see Table 2). It can be seen from the figure
that at the beginning the flux grows as $r^2$, which suggests that
the emission comes from an optical thick disk. Eventually (at
$r/R_0\simeq 40$), however, the shell becomes optically thin and the
flux tends to a constant value. The main contribution to
the flux comes from the optically thick region of the shell, and
therefore the emission from the optically thin region can be neglected.

  \begin{figure}[t]
   \centering
   \includegraphics[width=8cm]{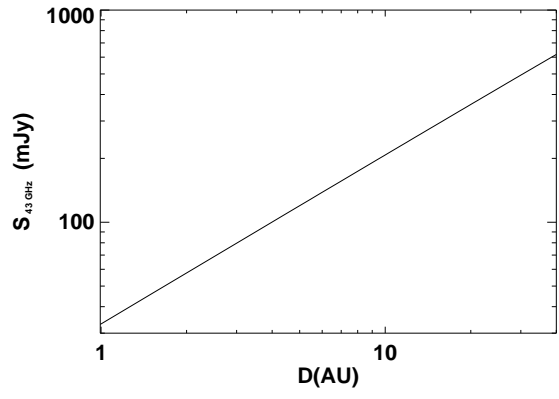}
      \caption{Predicted flux density at $S_{43 \, \rm{GHz}}$ from a binary
system as function of the binary separation $D$. We have adopted
the stellar wind parameters of Model B2 (see Table 2). From this example,
we found that the flux scales as $D^{0.79}$.}
         \label{f10}
   \end{figure}

Let $r_m$ be the transition impact parameter at which
$\tau_{s} (r_m)=$ 1. Since $EM_s\gg EM_w$ for lines of sight
with impact parameters $r\leq r_m$, we assume that the optical
depth is dominated by the shell. (For simplicity, we are not
considering those impact parameters $r\ll R_0$ for which
$EM_s\ll EM_w$; see Figure \ref{f7}.) It follows that the
optical depth $\tau (r_m)\simeq (2/\pi) A_{\nu} (R_0^2/r_m^5)\simeq 1$
(eq. [\ref{a16}]), and then the  size of the optically
thick region increases as $r_{m}\propto R_0^{2/5}$. 
Finally, since the total flux is dominated by the emission from
the optically thick region ($r\leq r_m$), one can assume that the
flux $S_{\nu}\propto r_{m}^2$. Consequently, we predict from our model
that the emission from the shocked layer scales as $S_{\nu}\propto D^{4/5}$.
Detailed calculations (as in the example shown in Fig. \ref{f10}),
which consider both the contribution of the optically thin region
to the emission from the shell and the contribution by the stellar
winds to the total optical depth, give a small deviation from this
prediction, that is, $S_{\nu}\propto D^{0.79}$.

\end{appendix}

\end{document}